\documentclass[conference]{IEEEtran}
\IEEEoverridecommandlockouts

\usepackage{cite}
\usepackage{amsmath,amssymb}
\usepackage{booktabs, multirow}
\usepackage{enumitem}
\usepackage{graphicx}
\usepackage{textcomp}
\usepackage{comment}
\usepackage{xcolor}
\usepackage[normalem]{ulem}
\usepackage{threeparttable}
\usepackage{listings}
\usepackage{subcaption}
\usepackage{url}
    
\definecolor{LightGray}{gray}{0.96}    
\definecolor{codegreen}{rgb}{0,0.6,0}
\definecolor{codegray}{rgb}{0.5,0.5,0.5}
\definecolor{codepurple}{rgb}{0.58,0,0.82}
\definecolor{backcolour}{rgb}{0.95,0.95,0.92}

\lstdefinestyle{mystyle}{
    backgroundcolor=\color{backcolour},   
    basicstyle=\ttfamily\footnotesize,
}
\begin{document}
\renewcommand\citepunct{, }

\newcommand{\gname}{GraphTango}

\title{
\gname{}: A Hybrid Representation Format for Efficient Streaming Graph Updates and Analysis
\\
}

\author{
\IEEEauthorblockN{Alif Ahmed, Farzana Ahmed Siddique, Kevin Skadron}
    \IEEEauthorblockA{University of Virginia}
    \IEEEauthorblockA{\{alifahmed,farzana,skadron\}@virginia.edu}
}

\maketitle

\thispagestyle{plain}
\pagestyle{plain}

\newcommand{\tnew}[2][c]{%
	\begin{tabular}[#1]{@{}c@{}}#2\end{tabular}}
\newcommand{\pluseq}{\mathrel{+}=}

\begin{abstract}



Streaming graph processing involves performing batched updates and analytics on a time-evolving graph. The underlying representation format largely determines the throughputs of these updates and analytics phases. Existing representation formats usually employ variations of hash tables or adjacency lists. However, a recent study showed that the adjacency-list-based approaches perform poorly on heavy-tailed graphs, and the hash-based approaches suffer on short-tailed graphs. We propose \gname{}, a hybrid representation format that provides excellent update and analytics throughput regardless of the graph's degree distribution. \gname{} dynamically switches among three different formats based on a vertex's degree: i) Low-degree vertices store the edges directly with the neighborhood metadata, confining accesses to a single cache line, ii)  Medium-degree vertices use adjacency lists, and iii) High-degree vertices use hash tables as well as adjacency lists. In this case, the adjacency list provides fast traversal during the analytics phase, while the hash table provides constant-time lookups during the update phase. We further optimized the performance by designing an open-addressing-based hash table that fully utilizes every fetched cache line. In addition, we developed a thread-local lock-free memory pool that allows fast growing/shrinking of the adjacency lists and hash tables in a multi-threaded environment. We evaluated \gname{} with the help of the SAGA-Bench framework and compared it with four other representation formats. On average, \gname{} provides 4.5x higher insertion throughput, 3.2x higher deletion throughput, and 1.1x higher analytics throughput over the \textit{next best} format. Furthermore, we integrated \gname{} with the state-of-the-art graph processing frameworks DZiG and RisGraph. Compared to the \textit{vanilla DZiG} and \textit{vanilla RisGraph}, [\textit{\gname{} + DZiG}] and [\textit{\gname{} + RisGraph}] reduces the average batch processing time by 2.3x and 1.5x, respectively.

\end{abstract}
\section{Introduction}










Streaming graph processing involves performing batched updates and analytics on a time-evolving graph. The update phase handles modifications to the graph topology (e.g., insertion/deletion of edges and nodes), while the analytics phase runs the necessary algorithms on the graph. This is a common scenario in many real-world graph applications such as social network analysis \cite{han2014chronos, cheng2012kineograph}, bioinformatics \cite{compeau2011apply,zerbino2008velvet}, recommendation systems \cite{grewal2018recservice, eksombatchai2018pixie}, routing and navigation \cite{park2019navigation}, knowledge discovery \cite{braun2016knowledge}, sensor networks \cite{borgman2007drowning}, etc. The focus of streaming graph processing is fundamentally different from static graph processing. \textit{Static graphs} are constructed only once, and the construction cost gets amortized over time. Therefore, the overall performance of a static graph processing framework is primarily determined by the analytics throughput. In the case of \textit{streaming graphs}, the graph topology can change very frequently. Hence, both update and analytics throughput is critical for streaming graphs \cite{basak2020saga}. 

The most common operation during the update phase is \textit{edge lookup}. The lookup is performed before insertion to avoid duplicate edges\footnote{In accordance with the prior works \cite{basak2020saga, ediger2012stinger, jaiyeoba2019graphtinker, iwabuchi2016towards}, edges are inserted only after a lookup to avoid duplicate edges.} and before deletion to find the location of the target edge. On the other hand, the most common operation during the analytics phase is the \textit{neighborhood traversal} of a given vertex. The performance of a streaming graph processing framework is critically dependent on how efficiently the graph storage format can support these lookup and traversal operations. Existing storage formats for streaming graphs usually employ variations of adjacency lists or hash tables \cite{iwabuchi2016towards, jaiyeoba2019graphtinker, basak2020saga, ediger2012stinger, mccrabb2021optimizing}. Approaches based on adjacency lists \cite{ediger2012stinger, basak2020saga} provide high update throughput on short-tailed graphs\footnote{\label{htail}Following prior work \cite{basak2020saga}, we define heavy/short-tailed graph with respect to an update batch: heavy-tailed graphs have high \textit{maximum degree} within a batch. Short-tail is the opposite.} but suffer in heavy-tailed graphs as it requires linear lookup through the edge array \cite{basak2020saga}.
On the other hand, hash-based approaches \cite{iwabuchi2016towards, jaiyeoba2019graphtinker} offer constant-time lookup, providing better update throughput on heavy-tailed graphs. However, they perform poorly on short-tailed graphs because the overhead of hash calculation and several random accesses becomes more expensive than conducting a simple linear search. Furthermore, edges are stored in hash tables relatively sparsely to mitigate collisions. As a result, edge traversal becomes inefficient and negatively impacts their analytics phase's throughput. None of the existing approaches can efficiently handle both short-tailed and heavy-tailed graphs.

This paper proposes \gname{}, a streaming graph representation format that provides excellent performance regardless of the graph's degree distribution. Our key idea is to adaptively switch the underlying data structure based on the vertex degree: i) \textit{Type1 vertex}: Low-degree vertices where the edges are stored within the same cache line as the neighborhood metadata. Update and edge traversal thus requires only one cache line access, unlike other approaches. ii) \textit{Type2 vertex}: Medium-degree vertices that store edges as adjacency lists. The degree is too high for this type to fit all edges in a cache line, but small enough so that linear search performs better than hashing. iii) \textit{Type3 vertex}: High-degree vertices that store edges as adjacency lists, along with hash tables storing indexes to the adjacency lists. In this case, the adjacency list provides optimal edge traversal during the analytics phase, while the hash table provides constant-time lookup during the update phase. The hash tables are not accessed during the analytics phase, avoiding any potential cache pollution. To improve the cache access pattern of the hash table, we designed an open-addressing-based hash table with double hashing that fully utilizes every fetched cache line. Our proposed hashing scheme minimizes cache line fetches and is especially beneficial if the hash tables do not fit into the last level cache (LLC), which is often the case for real-world graph workloads\footnote{Even with our smallest dataset of 5M edges, the LLC miss rate during the update phase is over 49\%, indicating that the working set size is larger than the LLC.}. With this hashing scheme, updates for Type3 vertices are performed with only three cache line accesses for more than 99.2\% of the cases. In addition, we developed a thread-local lock-free memory pool that allows fast growing and shrinking of the adjacency lists and hash tables in a multi-threaded environment.


We evaluated \gname{} by integrating it with the SAGA-Bench \cite{basak2020saga} benchmarking framework. SAGA-Bench integration ensures that all approaches use the same algorithm implementations via a common API. Therefore, any performance improvement comes purely from the data structure standpoint. SAGA-Bench comes with four representation formats: AdListShared, AdListChunked, Stinger \cite{ediger2012stinger}, and DegAwareRHH \cite{iwabuchi2016towards}, each of which is shown to excel in different algorithm and dataset combinations \cite{basak2020saga}. Details of these formats can be found in Section \ref{ssec:back_formats}. For update operations, \gname{} consistently performed best across all datasets. On average (maximum), \gname{} demonstrates 4.5x (6.6x) higher insertion throughput and 3.2x (5.0x) higher deletion throughput over the \textit{next best} approach. As for analytics, \gname{} offers 1.1x (1.6x) higher throughput than the \textit{next best} approach. Unlike prior approaches, \gname{} provides excellent update and analytics throughput for both short-tailed and heavy-tailed graphs.

Being a storage format, \gname{} is orthogonal to most full-fledged graph processing frameworks and can easily replace the underlying storage formats of those frameworks. To demonstrate, we integrated \gname{} with the state-of-the-art graph processing frameworks DZiG \cite{mariappan2021dzig} and RisGraph \cite{feng2021risgraph}. \textit{DZiG + \gname{}} reduced the overall batch processing runtime by 2.3x (5.2x) on average (maximum) compared to the original DZiG. \textit{RisGraph + \gname{}} reduced the overall batch processing runtime by 1.5x (1.9x) on average (maximum) compared to the original RisGraph.

\gname{} will be made available on GitHub, both as a standalone framework and as an integration with SAGA-Bench, DZiG, and RisGraph.
\section{Existing Representation Formats}
\label{sec:background}

\label{ssec:back_formats}

\begin{figure}[t]
	\centering
	\includegraphics[clip, trim={0cm 8.9cm 20.2cm 0.1cm}, scale=0.62]{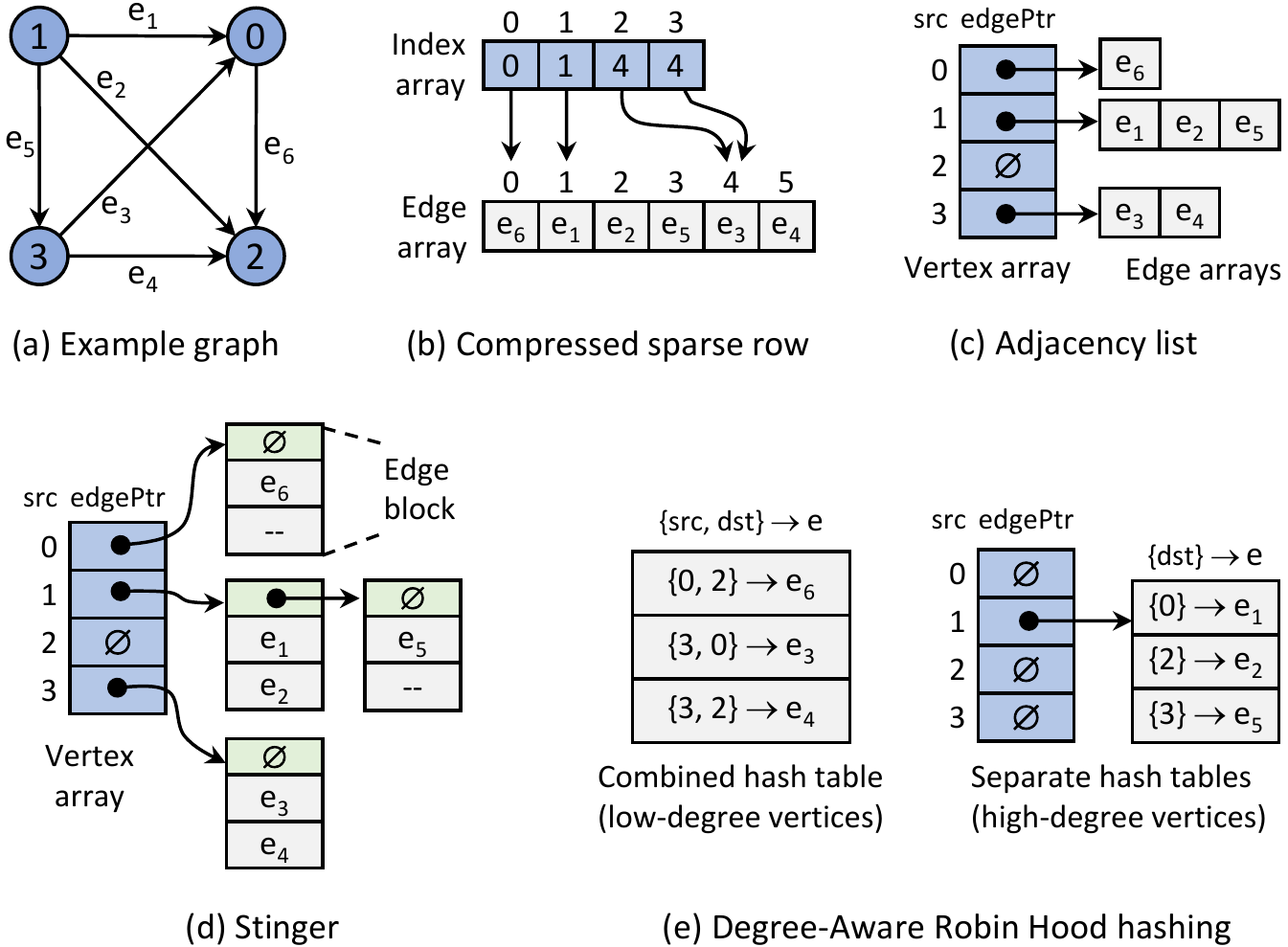}
	\caption{Example of different graph representation formats. Here, each edge $e$ is an $\{dst, prop\}$ tuple.}
	\label{fig:common_ds}
	\vspace{-0.5cm}
\end{figure}

\begin{figure*}[!t]
	\centering
	\includegraphics[clip, trim={0.3cm 9.3cm 1.3cm 0.4cm}, scale=0.55]{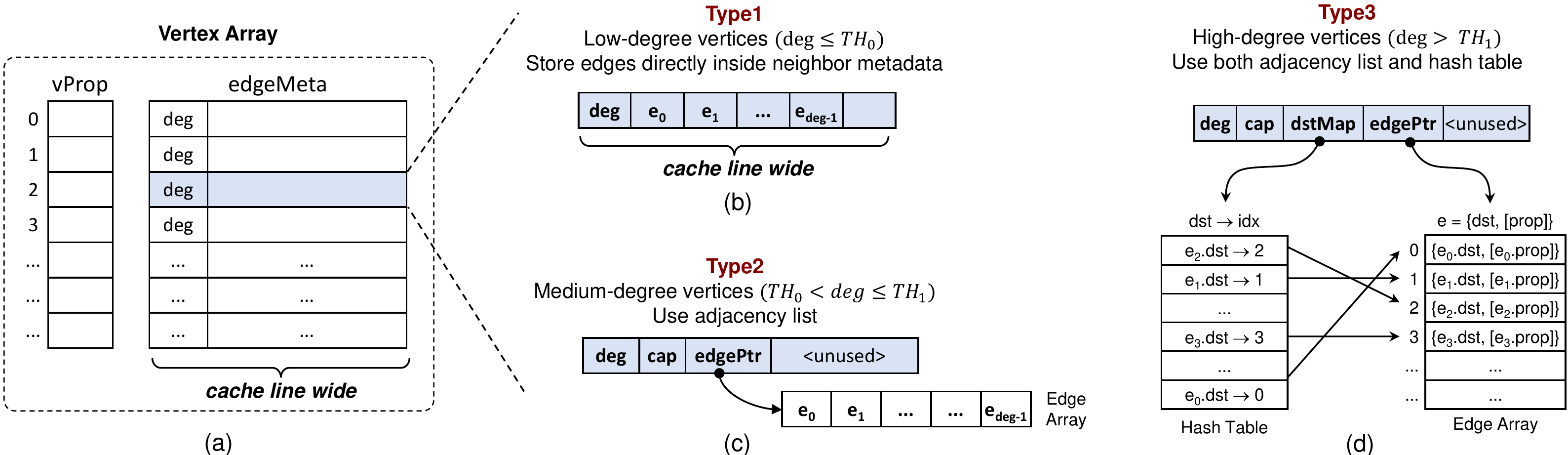}
	\caption{Proposed hybrid representation format of \gname{}.
	}
	\label{fig:graph_tango}
    \vspace{-0.5cm}
\end{figure*}

Figure \ref{fig:common_ds} illustrates how various graph representation formats store vertices and edges. While these examples store only the outgoing edges, the concept is also applicable if storing incoming edges.

\textbf{\textit{Compressed Sparse Row (CSR)}} is one of the most commonly used formats for \textit{static} graphs \cite{hu2021graphlily, ham2016graphicionado, sundaram2015graphmat, gui2019survey}. As shown in Figure \ref{fig:common_ds}(b), CSR organizes data in an \textit{edge array} and an \textit{index array}. Edges are stored in the edge array in ascending order - all edges of vertex $v_i$ appear before any edge of $v_{i+1}$. The index array stores the position of the first edge of every vertex. CSR is widely used for static graphs because it provides a compact representation, increasing spatial locality while traversing the graph. However, inserting or deleting an edge requires reconstructing both the edge array and the index array, making CSR unsuitable for dynamic graphs.

\textbf{\textit{Adjacency List}} stores the edges of every vertex in separate arrays (Figure \ref{fig:common_ds}(c)). A \textit{vertex array} stores the pointers to these edge arrays. \textbf{These edge arrays are assumed to be memory-contiguous (like \textit{std::vector}), rather than a linked list of edges.} This important distinction is used throughout the paper. As each edge array can grow/shrink independently, insertion and deletion operations only modifies the edge array of the corresponding vertex. This property makes adjacency lists a common choice for dynamic graph frameworks \cite{basak2020saga, mccrabb2021optimizing}. Another advantage of adjacency lists is that the edge traversal during the analytics phase has a sequential access pattern, leading to excellent analytics throughput for vertex-centric algorithms. The downside of adjacency lists is that the edges are not stored in any particular order within an edge array. Therefore, finding an edge requires a linear search through the corresponding edge array, leading to poor update throughput on high-degree vertices.

In adjacency-list-based approaches, parallel updates on multiple vertices are realized in two ways. The first scheme is the shared style multithreading (referred as \textit{\textbf{AdListShared}}), where the vertex array additionally contains a lock for every vertex. Any thread can process updates on any vertex by acquiring the corresponding lock first. This approach provides fine-grained parallelism. However, if most updates are targeted towards the \textit{same} vertex, it can cause lock contention and is often the case for heavy-tailed graphs. The alternative scheme groups source vertices into chunks and assign each chunk to a fixed thread (referred as \textit{\textbf{AdListChunked}}).
Chunked style multithreading is lock-free. However, it is prone to workload imbalance if the chunks have a high disparity in the number of edges they contain.

\textbf{\textit{Stinger}} \cite{ediger2012stinger} is an adjacency-list-based representation format. As illustrated in Figure \ref{fig:common_ds}(d), Stinger stores the edges as linked lists of \textit{edge blocks}. Each edge block can accommodate a fixed number of edges (default is 16). Parallelism in Stinger is achieved by acquiring locks on the edge blocks. The capacity of the edge blocks presents a trade-off between performance and storage requirements. Using smaller capacity edge blocks increase parallelism but makes graph traversal inefficient by increasing the amount of pointer-chasing accesses. On the other hand, larger blocks lead to many unused slots for low-degree vertices. Besides, like adjacency lists, Stinger also suffers from linear lookups on high-degree vertices, stagnating the update throughput.


\textbf{\textit{Degree-Aware Robin Hood Hashing (DegAwareRHH)}} \cite{iwabuchi2016towards} is a hash-based format. As shown in Figure \ref{fig:common_ds}(e), DegAwareRHH maintains two types of hash tables based on the vertex degree. Edges corresponding to low-degree vertices are stored in a combined hash table to improve data locality. On the other hand, each high-degree vertex maintains its own hash table. Both of these hash tables use Robin Hood hashing \cite{celis1985robin}, which minimizes probing distance. For parallelism, DegAwareRHH leverages chunked-style multithreading similar to AdListChunked. The constant time lookup enabled by the hash tables makes DegAwareRHH suitable for the update phases on heavy-tailed graphs. However, the sparse storage of edges in the hash table makes DegAwareRHH's edge traversal inefficient, negatively impacting the analytics throughput.

\section{\gname{} Data Structure}
\label{sec:data_structure}

Figure \ref{fig:graph_tango} gives an overview of the \gname{} data structure\footnote{The description assumes storing only outgoing edges for clarity. In our implementation, we stored both incoming and outgoing edges for directed graphs.}. \gname{} organizes the vertex data in two arrays: one for storing the vertex properties ($vProp$) and the other for storing neighborhood metadata of the vertex ($edgeMeta$). These arrays are indexed using vertex id. Neighbors of each vertex are stored as an $e_x = \{dst, [prop]\}$ tuple, where $e_x.dst$ is the destination vertex id, and $e_x.prop$ is an optional edge property (e.g., the weight of the edge). 

The $edgeMeta$ array is aligned to a page boundary\footnote{To clarify, only the $edgeMeta$ array itself is page boundary aligned, not the edge arrays or hash tables it may point to.}, and each element of the array is of cache line size. Therefore, accessing any field of $edgeMeta[i]$ will bring the rest of the fields into the cache. The $deg$ field holds the current degree of the corresponding vertex. Depending on the degree, a vertex will fall into one of the following three categories:

\textbf{\textit{Type1 Vertex:}} These are low-degree vertices with $deg \leq TH_0$. As illustrated in Figure \ref{fig:graph_tango}(b), we store the edges directly with the metadata for Type1 vertices. The threshold $TH_0$ denotes the number of edges that can fit inside the metadata and is defined as:
$$TH_0 = \left\lfloor{ \frac{ CACHE\_LINE\_SIZE - sizeof(deg)}{sizeof(e)}}\right\rfloor $$

For example, $TH_0 = 7$ for a typical cache line size of 64 bytes and edges of 8 bytes. The advantage of storing edges with metadata is that all edges are brought into the cache as soon as we access the vertex during the update or analytics phase. When searching for a specific edge, we need to do a linear search. However, the search is extremely fast, as all accesses will be cache hits.

\textbf{\textit{Type2 Vertex:}} These are medium-degree vertices with $TH_0 < deg \leq TH_1$, where $TH_1$ is a user-configurable threshold. Edges for this type of vertices are stored in adjacency lists, as shown in Figure \ref{fig:graph_tango}(c). To support adjacency lists, $edgeMeta$ additionally maintains the current capacity ($cap$) and a pointer to its edge array ($edgePtr$).

Like Type1 vertices, Type2 also requires a linear search when looking for a specific edge. As the linear search on the edge array is prefetcher-friendly and has good spatial locality, it offers better performance than hash-based search up to a certain point (i.e., tuned using the $TH_1$ threshold). However, the linear nature of the search becomes a performance bottleneck for higher-degree vertices.
Hash-based search is preferable in such cases, as explained below.


\textbf{\textit{Type3 Vertex:}} These are high-degree vertices with $deg > TH_1$. Figure \ref{fig:graph_tango}(d) illustrates this scenario. 
Here, we maintain \textit{both an adjacency list and a hash table for each Type3 vertex}. 
The hash table maps an edge's destination vertex id ($e_x.dst$) with its location in the corresponding adjacency list. Maintaining both hash table and adjacency list comes with the following benefits: i) The hash table enables constant-time lookups during the \textit{update phase}. ii) The adjacency list provides fast and efficient traversal during the \textit{analytics phase}. Prior hash-based approaches suffer from low analytics throughput due to inefficient edge traversal \cite{basak2020saga}. \gname{} is free of this issue because it uses only the adjacency lists for edge traversal and does not require accessing the hash tables during the entirety of the analytics phase.





\section{\gname{} Basic Operations}
\label{sec:basic_op}

\begin{table}[b]
\centering
\caption{Vertex type switching steps for insertion/deletions}
\label{tab:overhead}
\renewcommand{\arraystretch}{1.1}
\scriptsize
\begin{subtable}{0.5\textwidth}
\setlength\tabcolsep{2.8pt}
\vspace{-0.1cm}
\caption{Insertions triggering type switch or capacity doubling}
\vspace{-0.1cm}
\begin{tabular}{cccccc}
\toprule
\tnew{Direction}&New capacity & \tnew{Alloc new\\edge array} & \tnew{Edge copy\\size} & \tnew{Dealloc old\\ edge array} & Rehash \\
\midrule
Type1 $\rightarrow$ Type2 & nextPow2($TH_0$) & $\checkmark$ & deg (=$TH_0$) & X & X \\
Type2 $\rightarrow$ Type2 &cap * 2 & $\checkmark$ & deg & $\checkmark$ & X \\
Type2 $\rightarrow$ Type3 &cap * 2 & $\checkmark$ & deg (=$TH_1$) & $\checkmark$  & $\checkmark$ \\
Type3 $\rightarrow$ Type3 &cap * 2 & $\checkmark$ & deg & $\checkmark$ & $\checkmark$ \\
\bottomrule
\end{tabular}
\end{subtable}
\hfill
\vspace{0.2cm}
\begin{subtable}{0.5\textwidth}
\setlength\tabcolsep{3.4pt}
\caption{Deletions triggering type switch or capacity halving}
\vspace{-0.1cm}
\begin{tabular}{cccccc}
\toprule
\tnew{Direction}&New capacity & \tnew{Alloc new\\edge array} & \tnew{Edge copy\\size} & \tnew{Dealloc old\\ edge array} & Rehash \\
\midrule
Type3 $\rightarrow$ Type3 & cap / 2 & $\checkmark$ & deg & $\checkmark$ & $\checkmark$ \\
Type3 $\rightarrow$ Type2 & cap / 2 & $\checkmark$ & deg (=$TH_1$) & $\checkmark$ & X \\
Type2 $\rightarrow$ Type2 & cap / 2 & $\checkmark$ & deg & $\checkmark$  & X \\
Type2 $\rightarrow$ Type1 & $TH_0$ & X & deg (=$TH_0$)& $\checkmark$ & X \\
\bottomrule
\end{tabular}
\end{subtable}
\end{table}

\subsection{Edge Insertion}
The edge insertion procedure is as follows: (i) Retrieve the edge metadata - $edgeMeta[srcId]$. (ii) If the current $deg$ reaches the current capacity
, we double the capacity. The exact steps for capacity doubling will depend upon the current and new type, as demonstrated in Table \ref{tab:overhead}(a). In general, capacity doubling involves allocating memory for the larger edge array, copying current edges to the new edge array, and freeing the old array. For Type3, the hash table is also rehashed. The amortized cost of capacity doubling is O(1) \cite{cormen2009introduction}.  (iii) Search for a duplicate edge using $dst$. As mentioned earlier, for Type1 and Type2, it will involve doing a linear search, and for Type3, the search will be performed using the hash table. (iv-A) If the edge is found, update the property and return. (iv-B) If the edge is \textit{not} found, add the edge at the end of the edge array and increment $deg$. For Type3, we also create an entry in the hash table pointing to the location.


\subsection{Edge Deletion}
\label{ssec:edge_del}

The edge deletion procedure is as follows: (i) Retrieve the edge metadata - $edgeMeta[srcId]$. (ii) Search for existing edge using $dst$. (iii-A) If the edge is \textit{not} found, return. (iii-B) If the edge is found, delete the entry from the edge array and hash table (for Type3) and decrement $deg$. We do a \textit{\textbf{compaction step}} here to fill the gap. It involves moving the last entry of the edge array to the deleted entry's position and updating the corresponding hash table record. The compaction step is simple and is of constant time complexity. (iv) If the $deg$ becomes 1/4th of the capacity, we halve the capacity. The steps for capacity halving is given in Table \ref{tab:overhead}(b). Similar to the capacity doubling during insertion, the amortized cost of capacity halving is also O(1) \cite{cormen2009introduction}.

\subsection{Edge Traversal}
As we store the edges in consecutive memory for all three vertex types\footnote{Even after deletions, our compaction step ensures that all valid edges of a vertex are stored in consecutive memory.}, the edge traversal API simply returns a cursor (i.e., position of the iterator) for indexing to the: i) $edgeMeta[vid]$ for Type1 vertices, or ii) $edgeMeta[vid].edgePtr$ for Type2/Type3 vertices. \gname{}'s traversal mechanism is essentially the same as an adjacency list for Type2/Type3 vertices. As for Type1, \gname{} has a better access pattern as it requires one less indirection.



\section{Optimizing \gname{}}
\subsection{Cache-Friendly Hashing Scheme}

The hash table used by the Type3 vertices can be realized in several ways. The most convenient approach is to use \textit{std::unordered\_map}. Unfortunately, this approach is not ideal for our purpose because the C++ standard \cite{c2020standard} effectively limits the collision resolution of \textit{std::unordered\_map} to separate chaining\footnote{This constraint is a side effect of mandating \textit{pointer stability}, which means that an iterator must remain valid upon inserting or deleting elements.}. With separate chaining, the hash table is constructed as an array of buckets. Each bucket points to a linked list of colliding elements (i.e., keys that hashed to the same bucket). The issue with separate chaining is that it involves multiple random accesses - one to access the bucket and one or more for traversing the linked list. Each of these random accesses is a potential cache miss if the hash table does not fit into the cache. An alternative to separate chaining is open addressing, where all elements are stored in the hash table itself, eliminating the need for linked lists traversals. Prior hash-based graph representation formats \cite{iwabuchi2016towards, jaiyeoba2019graphtinker} leveraged open-addressing-based Robin Hood hashing \cite{celis1985robin} that minimizes probing distance. For \gname{}, we designed a more cache-friendly open-addressing-based hash table that minimizes the number of cache line accesses, making it especially suitable for real-world graph workloads where the hash table is unlikely to fit into the cache.

\begin{figure}[t]
	\centering
	\includegraphics[clip, trim={0.5cm 9.2cm 19.8cm 0.1cm}, scale=0.62]{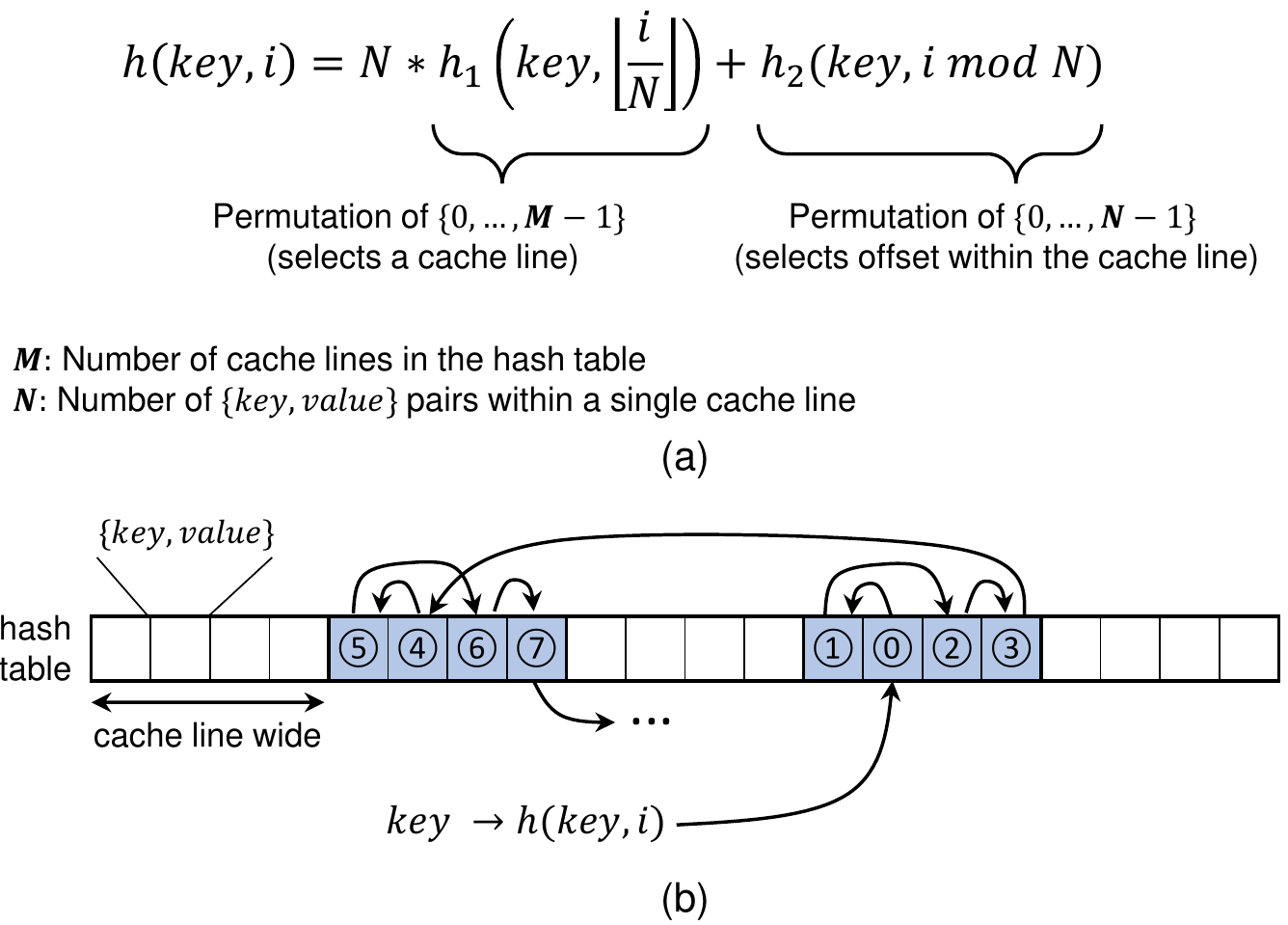}
	\caption{Proposed hashing scheme. (a) Hash function to determine the index to the hash table. (b) An example probing sequence for $M=5$ and $N=4$.}
	\label{fig:hash_scheme}
\end{figure}


The key idea of our hashing scheme is to limit the probes within a single cache line until it is fully searched, before moving onto a different cache line.
Figure \ref{fig:hash_scheme} illustrates this hashing scheme. 
The hash table itself is composed of an array of $\{key,value\}$ pairs. The index of the $i$-th probe to the hash table is given by the following hash function:
\begin{equation*}
    h(key,i)=N \cdot h_1\left(key, \left\lfloor \frac{i}{N} \right\rfloor\right) + h_2(key, i~mod~N)
\end{equation*}

Here, $N$ is the number of $\{key,value\}$ pairs that can fit within a single cache line. The purpose of $h_1\left(key, \left\lfloor \frac{i}{N} \right\rfloor\right)$ is to select a cache line for probing and returns the base index of that selected cache line. Note that the $\left\lfloor \frac{i}{N} \right\rfloor$ parameter remains the same for every $N$ consecutive probes, thereby selecting the same cache line. As $h_1()$ should eventually explore all cache lines in the hash table, it must be a permutation of $\{0,1,...,M-1\}$, where $M$ is the number of cache lines in the hash table. On the other hand, $h_2(key, i~mod~N)$ determines the offset within the cache line and must be a permutation of $\{0,1,...,N-1\}$. Any hash function conforming to this permutation requirement can be used to implement $h_1()$ and $h_2()$. 
In \gname{}, we used double hashing for $h_1()$ to avoid primary/secondary clustering. $h_2()$ uses linear probing to make hash computation simpler. 
Our hash function is very cheap to compute, with the reference implementation having two multiplications and eight other simple arithmetic/logical instructions. This is because we ensure that both $N$ and $M$ are powers-of-two, converting expensive modulus and division operations to simple shifts. 
Further optimization is possible by leveraging SIMD instructions to do a parallel comparison on all entries mapped to the same cache line. However, as discussed later, \gname{} demonstrates short probing distance, making iterative comparison just as performant.

Insertions and deletions to the proposed hash table are similar to other open-addressing-based hash tables. Each location of the hash table can contain either: i) a valid $\{key,value\}$ pair, or ii) an \textit{empty} marker, or iii) a \textit{deleted} marker (i.e., tombstone). We used two reserved values as the empty and deleted marker instead of using dedicated tag storage. During both insertion and deletion, the table is probed (using the hash function) until the $key$ or an empty marker is found. If the $key$ is found: i) For insertion, the corresponding $value$ is updated. ii) For deletion, the entry is marked as deleted. Instead of $key$, if an empty marker is found: i) No action is required for deletion. ii) For insertion, the $\{key, value\}$ pair is inserted to the location of the first encountered delete marker, or to the current location if no delete marker was encountered.

When using with \gname{}, the hash tables' initial capacity is set to twice the capacity of the corresponding adjacency lists. Upon inserting/deleting edges, both the hash tables and the adjacency lists can grow/shrink in size (see Section \ref{sec:basic_op}), but the capacity ratio always remains 2. This property sets the maximum load factor ($\alpha$) of the hash table to $0.5$. Assuming uniform hashing\footnote{Double hashing can demonstrate performance very close to the ideal scenario of uniform hashing \cite{cormen2009introduction, knuth1973art}.}, the theoretical average probing distance is: i) $\frac{1}{1-\alpha} = 2$ for an unsuccessful search. This is often the case for edge insertions in the absence of duplicates. ii) $\frac{1}{\alpha}ln\frac{1}{1-\alpha} = 1.39$ for a successful search (e.g., deleting existing edges). In \gname{}, we can fit eight $\{key,value\}$ pairs within a single cache line (i.e., $N=8$). As a result, the hash table needs to access only one cache line as long as the probing distance remains $\leq 8$, which provides a large slack over the theoretical average probing distances. We empirically observed the same trend with our graph datasets, where over 99.2\% of the insertions had a probing distance $\leq 8$. Therefore, almost all edge insertion operations for \textit{Type3} vertices require only three cache line accesses: i) one for retrieving $edgeMeta[srcId]$ metadata that contains hash table and adjacency list pointers, ii) one for searching the hash table, and iii) one for indexing to the adjacency list.




\subsection{Memory Allocation Scheme}
As discussed in Section \ref{sec:basic_op}, \gname{} requires frequent growing/shrinking of adjacency lists and hash tables. Calling $malloc()$/$free()$ in every such instance can cause high runtime overhead and memory fragmentation. We avoid this issue by designing a fast thread-local lock-free memory pool that supports O(1) allocation and deallocation.

\begin{figure}[h]
	\centering
	\includegraphics[clip, trim={0.2cm 4cm 21.6cm 0.0cm}, scale=0.55]{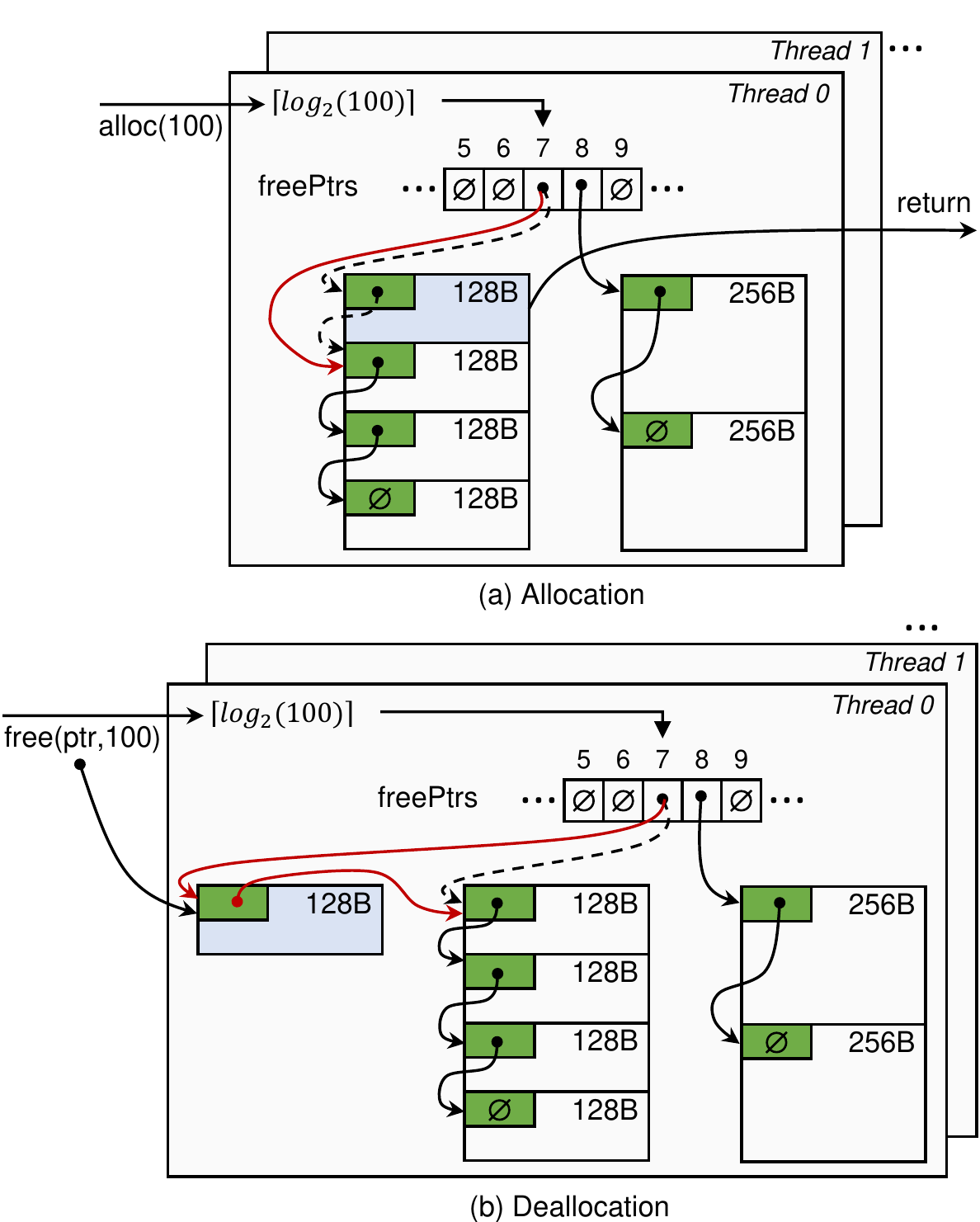}
	\caption{Allocation and deallocation on the memory pool. 
	Deleted pointers are shown by dashed lines and the modified pointers by red lines.}
	\label{fig:mem_pool}
\end{figure}

Figure \ref{fig:mem_pool} illustrates the data structure of the memory pool. This memory pool allocates chunks in power-of-two  sizes. Individual linked lists of available chunks are maintained for each valid size. The heads of the linked lists are stored in the $freePtrs$ array. The first 8 bytes of each chunk (highlighted green) hold the pointer to the next free chunk of the same size. This way, no extra storage beside the $freePtrs$ array is required to hold the pointers. However, it limits the minimum chunk size to 8 bytes in a 64-bit machine.

\textit{\textbf{Allocation}} steps are shown in Figure \ref{fig:mem_pool}(a). For an allocation request of $sz$ bytes, the pool will return a chunk of size $newSz = 2^k$, where $k=\lceil log_2(sz)\rceil$ (i.e., nearest power of two that is $\ge sz$). The allocation proceeds as follows: i) Find the first available free chunk of $newSz$. This is simply given by $ret = freePtrs[k]$. ii) If $ret$ is not a null pointer, then it points to a free chunk. In this case, we update $freePtrs[k]$ to point to the next free chunk, and return $ret$. iii) If $ret$ is a null pointer, this indicates no chunk of the requested size is available. In this case, a large memory block is allocated (of size $max(4MB, newSz)$ and aligned to the page boundary) and then split into a linked list of $newSz$ byte chunks. $freePtrs[k]$ is set to point to the first chunk. At this point, we have free chunks of $newSz$. Therefore, repeating step (ii) will complete the allocation. Note that, once allocated, the full chunk can be used to store data, including the space initially used to hold the pointer to the next chunk.

\textit{\textbf{Deallocation}} steps are shown in Figure \ref{fig:mem_pool}(b). Unlike the standard $free(ptr)$, we provide the size of the allocated chunk as an additional parameter - $free(ptr, sz)$. Using the $sz$ parameter, we can directly index to the $freePtrs$ array and add $ptr$ as a free chunk, as shown in Figure \ref{fig:mem_pool}(b). 

An advantage of the proposed memory pool is that the most recently deallocated chunk will be allocated first, thereby being more likely to reside in the cache. Furthermore, each thread maintains its own $freePtrs$ array. As a result, no lock is required when multiple threads are trying to allocate/deallocate simultaneously.
A minor downside is that one thread cannot allocate free chunks from another thread's pool. We found it to be of little consequence in practice because the maximum amount of unused space per thread is O(blockSize). 
Also, note that the $\lceil log_2(sz)\rceil$ calculation used to index $freePtrs$ is very cheap to perform. It only requires \textit{count leading zero} $(clz)$ and shift instructions.

\subsection{Parallelization}

For \gname{}, we experimented with both shared and chunked style multithreading. We decided to settle for chunked style multithreading as it demonstrated slightly better throughput on our datasets. As we are using chunked style multithreading, operations concerning any vertices within a partition is mapped to a fixed worker thread. As discussed before, this method eliminates the need of performing atomic operations and requires minimum synchronization, but may suffer from workload imbalance. Exploring advanced load-balancing techniques is left as a potential future work.
Another related concern is false sharing, which might occur when multiple threads simultaneously modify a shared data structure. In the case of \gname{}, false-sharing-prone data structures are the vertex property array and the active frontier array. We avoid false sharing by using a partition size multiple of the cache line size (e.g., 512). It ensures that no two partitions' data share the same cache line.

\subsection{Determining the $TH_1$ Threshold}

Unlike the $TH_0$ threshold, which is fixed for a given cache line size and edge element size, the $TH_1$ threshold is flexible and has a moderate impact on performance and memory usage (Section \ref{ssec:th1}). As mentioned before, $TH_1$ should be set to a value for which $O(TH_1)$ linear search through the edge array is likely to perform better than $O(1)$ hash table lookup. The following equation provides an estimate and can be used as a rule of thumb for selecting $TH_1$:
\begin{equation}
\label{eq:th1}
TH_1=2^{\lceil log_2(3\times edgesPerCacheLine)\rceil}
\end{equation}

This equation sets $TH_1$ to a value roughly corresponding to four cache line accesses for Type2 vertices. This is slightly above the three cache line accesses of Type3 vertices, as Type2 vertices have favorable sequential access patterns and do not incur hash calculation overheads. As an alternative, we provide a microbenchmark program (\textit{graph dataset agnostic}) with \gname{} that empirically finds a suitable $TH_1$ threshold.


\section{Evaluation}
\subsection{Experimental Setup}
\label{ssec:setup}
\begin{figure*}
	\centering
	\includegraphics[clip, trim={0.1cm 10.3cm 2.8cm 0cm}, scale=0.57]{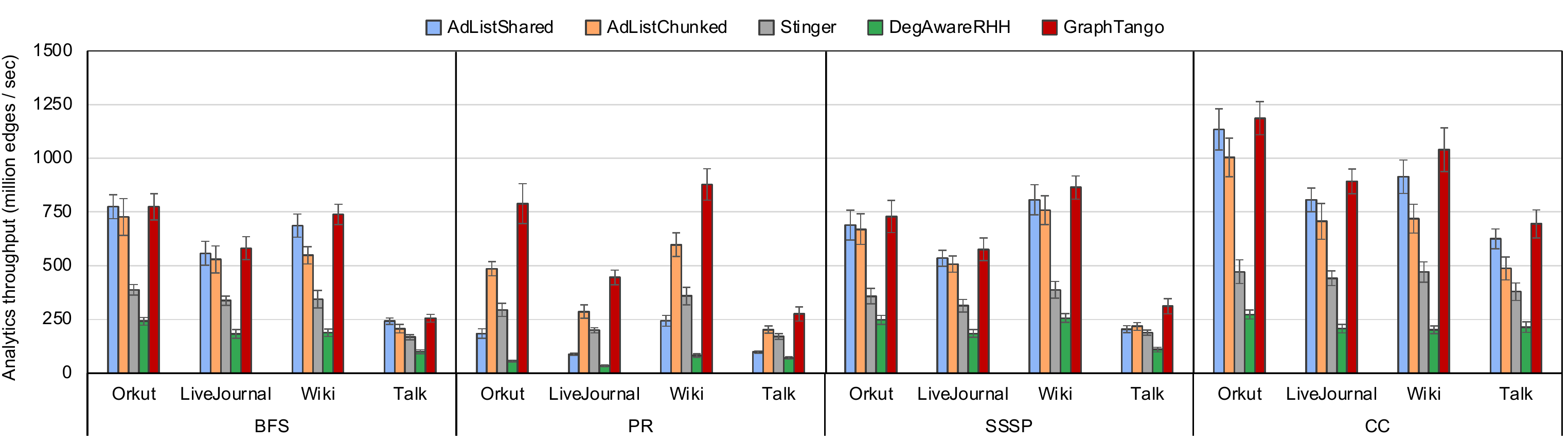}
	\caption{Comparison of the analytics throughputs. \textit{Higher is better}.
	}
	\label{fig:analytics_th}
\end{figure*}

\textbf{\textit{A) Platform:}} The experiments are conducted on an AMD Ryzen 3900x @ 3.8GHz machine with 12 physical cores, 64MB of LLC, and 32GB of DDR4 DRAM. Hyper-threading and turbo-boost were disabled for better reproducibility. All experiments are performed with 12 cores.

\textbf{\textit{B) Implementation:}}
We evaluated \gname{} by integrating it with the SAGA-Bench \cite{basak2020saga} benchmarking framework.
SAGA-Bench comes with four representation formats - AdListShared, AdListChunked, Stinger \cite{ediger2012stinger}, and DegAwareRHH \cite{iwabuchi2016towards}. Details of these formats can be found in Section \ref{ssec:back_formats}. 
SAGA-Bench integration facilitates fair comparison, because all approaches must use the exact same algorithm implementations through a common API\footnote{We used the official integration of SAGA-Bench from \cite{sagagit} even if the original author's source code is publicly available (e.g., Stinger \cite{stingergit}).}. The source code is compiled with gcc-9.3.0 and -03 flag.

Both vertex id and edge property are considered to be of 64-bits size. Therefore, \gname{} has $TH_0 = 7$ for unweighted graphs and $TH_0 = 3$ for weighted graphs. $TH_1$ is set to 32 following the tuning carried out in Section \ref{ssec:th1}. This $TH_1$ value also matches the value provided by Equation \ref{eq:th1}.

\textbf{\textit{C) Profiling Methodology:}}
Graph datasets are first randomly shuffled to break any existing ordering of edges. This is done to reflect the realistic scenario where edge updates are unlikely to occur in any pre-defined order. The shuffled dataset is inserted in batches of 1M edges until the full graph is built  and then deleted\footnote{The vanilla SAGA-Bench does not support edge deletions. We added deletion support for all representation formats by closely following the corresponding papers or from their source code if available.} in batches of 1M edges until no edges are left to delete. This batch size is similar to prior works \cite{jaiyeoba2019graphtinker, basak2020saga}. \textbf{\textbf{Analytics is performed on the graph after every batch of insertions and deletions. Reported throughputs are the geometric mean of the per-batch throughputs.}} \gname{} dynamically switches between vertex types as edges are inserted/deleted. As discussed in Section \ref{sec:basic_op}, this switching may involve memory allocation/deallocation, copying, or rehashing. \textbf{\textbf{This switching overhead is included in the reported results.}}


\textbf{\textit{D) Datasets:}}
We have used four real-world datasets in our experiments: Orkut, LiveJournal, Wiki-topcats (referred as Wiki), and Wiki-talk (referred as Talk). Orkut and LiveJournal are online social media networks, Wiki is a dataset of Wikipedia hyperlinks, and Talk is the Wikipedia communications network. These datasets are part of the SNAP dataset collection \cite{leskovec2014snap}. All these datasets are directed except for Orkut. Properties of these datasets are given in Table \ref{tab:datasets}. Orkut and LiveJournal have a much lower per-batch maximum degree compared to Wiki and Talk. Consequently, \textbf{Orkut and LiveJournal are characterized as short-tailed graphs while Wiki and Talk are heavy-tailed graphs.}

\begin{table}[t]
    \centering
    \footnotesize
    \caption{Evaluated Datasets}
    \renewcommand{\arraystretch}{1.2}
    \setlength\tabcolsep{3.2pt}
    \begin{threeparttable}
    \begin{tabular}{l|c|c|c|c|c|c|c}
        \toprule
        \multirow{2}{*}{Dataset} & Vertices & Edges & \multicolumn{2}{c|}{Max degree\tnote{1}} & \multicolumn{3}{c}{Vertex mapping\tnote{2}}\\
        & (million) & (million) & in & out & Type1 & Type2 & Type3 \\
        \midrule
        Orkut & 3.0 & 117.2 & 329 & 329 & 27.2\% & 38.6\% & 34.2\% \\
        LiveJournal & 4.8 & 69.0 & 237 & 332 & 63.0\% & 26.0\% & 11.0\% \\

        Wiki & 1.8 & 28.5 & 8,504 & 154 & 57.8\% & 33.9\% & 8.3\% \\
        Talk & 2.4 & 5.0 & 665 & 20,088 & 98.3\% & 1.2\% & 0.5\%\\
        \bottomrule
    \end{tabular}
    \begin{tablenotes}
        \scriptsize
        \item[1] Per-batch maximum degree with batch size of 1 million edges
        \item[2] For $TH_0=7$ and $TH_1=32$
    \end{tablenotes}
    \end{threeparttable}
    \label{tab:datasets}
\end{table}

\textbf{\textit{E) Algorithms:}}
We used four algorithms in our experiments: i) Breath-First Search (BFS), ii) Page Rank (PR), iii) Single-Source Shortest Path (SSSP), and iv) Connected Components (CC). Vertex centric incremental compute model is used for these algorithms, where the computation is constrained within the region affected by the update phase instead of the whole graph. The implementations of these algorithms are directly taken from SAGA-Bench without any modification. 

\subsection{Analytics and Update Performance}
\label{ssec:expt_perf}

\textbf{\textit{A) Analytics Throughput:}}
Figure \ref{fig:analytics_th} shows the analytics throughput of the representation formats. As mentioned in Section \ref{ssec:setup}(C), the analytics phase is conducted multiple times as we gradually build the graph. Reported values are the geometric mean of per-batch throughputs. \gname{} outperforms other approaches in every dataset and algorithm combinations. Compared to the \textit{next best} approach (i.e., AdListShared for BFS, SSSP, CC and AdListChunked for PR), \gname{} provides an avg (max) speedup of 1.1x (1.6x).
As all these approaches are using the exact same algorithm implementation, their relative performance is primarily determined by their edge traversal efficiency. Adjacency-list-based approaches perform well in this regard, because their edge traversal consists of mostly sequential accesses. \gname{} also uses adjacency lists for medium- and high-degree vertices (Type2 and Type3). For low-degree vertices (Type1), \gname{} has a better access pattern, as it requires one less indirection (i.e., does not need pointer chasing to find the corresponding edge array), thereby offering higher throughput. Stinger, despite using coarse-grained adjacency lists, suffers due to additional pointer chasing between edge blocks. Overall, \gname{} provides an avg (max) speedup of 1.8x (5.1x) over AdListShared, 1.3x (1.6x) over AdListChunked, 2.0x (2.7x) over Stinger, and 5.2x (14.0x) over DegAwareRHH.

  

\begin{figure}[h]
	\centering
	\includegraphics[clip, trim={0.1cm 5.5cm 21.3cm 0.1cm}, scale=0.60]{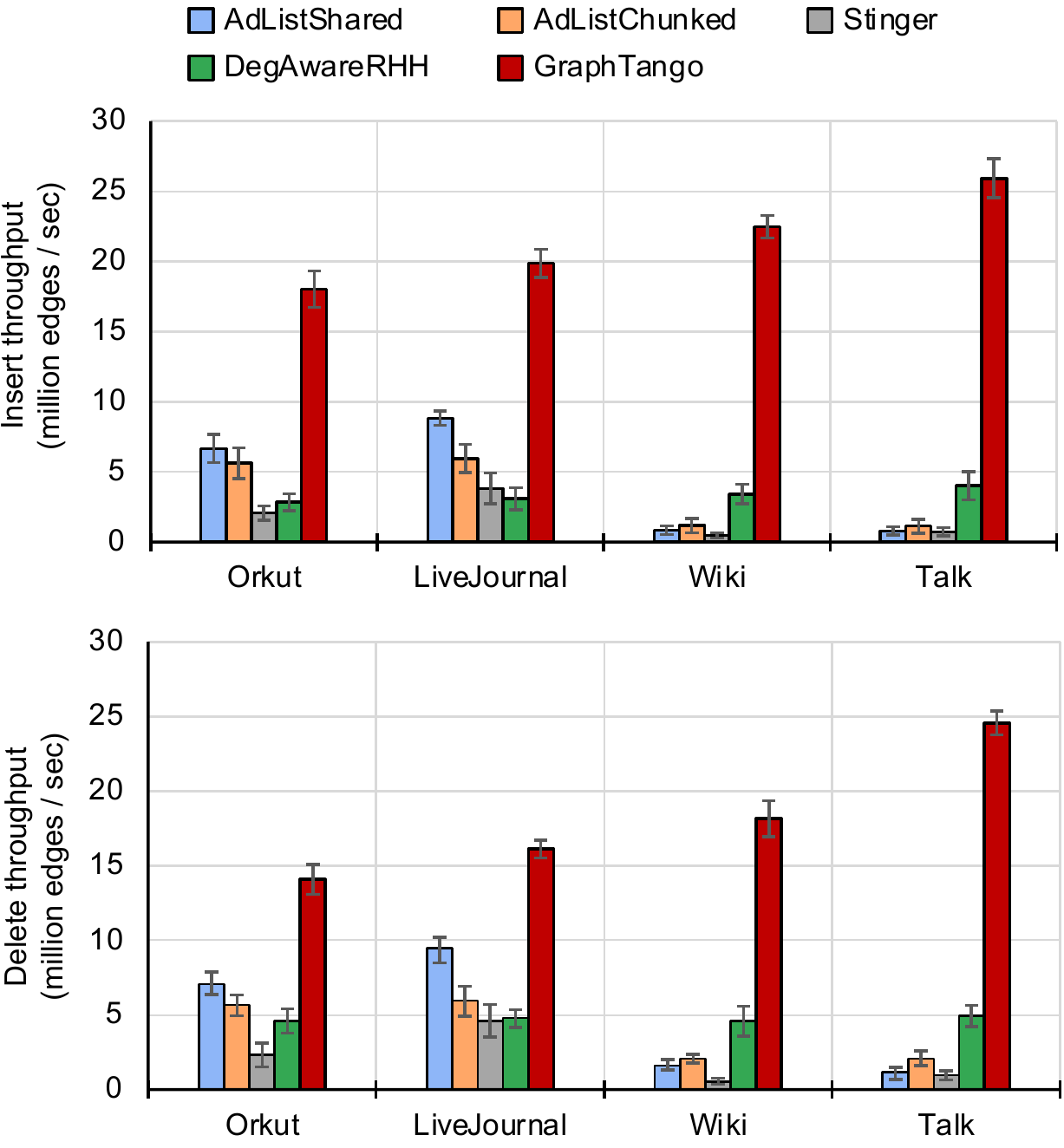}
	\caption{Comparison of update throughputs. \textit{Higher is better}.}
	\label{fig:ins_del_all}
\end{figure}

\textbf{\textit{B) Update Throughput:}}
Figure \ref{fig:ins_del_all} shows the update (edge insertion and deletion) throughput. Note that the updates are interleaved with analytics phases (see Section \ref{ssec:setup}.C). The algorithm choice of the analytics phase has little impact on the update throughput, and the reported values are the average across the four algorithms. Here, \gname{} outperforms other approaches by a large margin. Adjacency-list-based approaches perform well on short-tailed graphs. On these graphs, \gname{} provides an avg (max) speedup of 2.5x (2.7x) over the next best approach AdListShared. On the other hand, hash-based DegAwareRHH performs best on the heavy-tailed graphs. Interestingly, AdListShared performed even worse than AdListChunked for heavy-tailed graphs. This is due to the lock contention of shared-style multithreading on AdListShared. On heavy-tailed graphs, \gname{} provides an avg (max) speedup of 6.5x (6.6x) over the next best approach DegAwareRHH.
Notably, other approaches are suitable for either short- or heavy-tailed graphs. \gname{}'s hybrid nature makes it consistently the best-performing irrespective of the graph's degree distribution.

\begin{table}[b]
    \centering
    \caption{Average Memory Usage (Bytes Per Edge)}
    \label{tab:mem_req}
    \renewcommand{\arraystretch}{1.2}
    \setlength\tabcolsep{5pt}
    \scriptsize
    \begin{tabular}{l|c|c|c|c|c}
        \toprule
        \multirow{2}{*}{Dataset} & AdList- & AdList- & \multirow{2}{*}{Stinger} & DegAware & \multirow{2}{*}{\gname{}} \\
        & Shared & Chunked & & RHH &  \\
        \midrule
        Orkut &8.8 &12.0 &32.7 &43.8 &33.6 \\
        LiveJournal &10.2 &12.9 &48.9 &57.0 &34.6 \\
        Wiki &10.0 &12.7 &44.1 &62.9 &34.4 \\
        Talk &23.3 &21.9 &230.2 &74.6 &116.9 \\
        \bottomrule
    \end{tabular}
\end{table}

\subsection{Memory Usage}

Table \ref{tab:mem_req} shows the average memory usage per edge. The AdListShared and AdListChunked are most efficient in terms of memory usage. Compared to AdListShared - Stinger, DegAwareRHH, and \gname{} require 5.7x, 5.0x, and 3.9x more memory on average, respectively. 
For DegAwareRHH, the high memory usage is caused by: i) Sparse storage of edges in hash tables (to reduce collision), and ii) Robin Hood hashing mechanism that requires storing the probe distance for each entry. On the other hand, Stinger and \gname{} have a relatively high initial capacity (16 for Stinger and $TH_0$ for \gname{}) that remains mostly unused for low-degree vertices. This scenario is especially noticeable for the Talk dataset, where more than 96\% of the vertices have a degree $\leq 3$, leading to high memory usage. One way to reduce memory usage of \gname{} is to increase the $TH_1$ threshold, as described in Section \ref{ssec:th1}.


\begin{figure}[t]
	\centering
	\includegraphics[clip, trim={0.1cm 11.4cm 20.2cm 0.3cm}, scale=0.62]{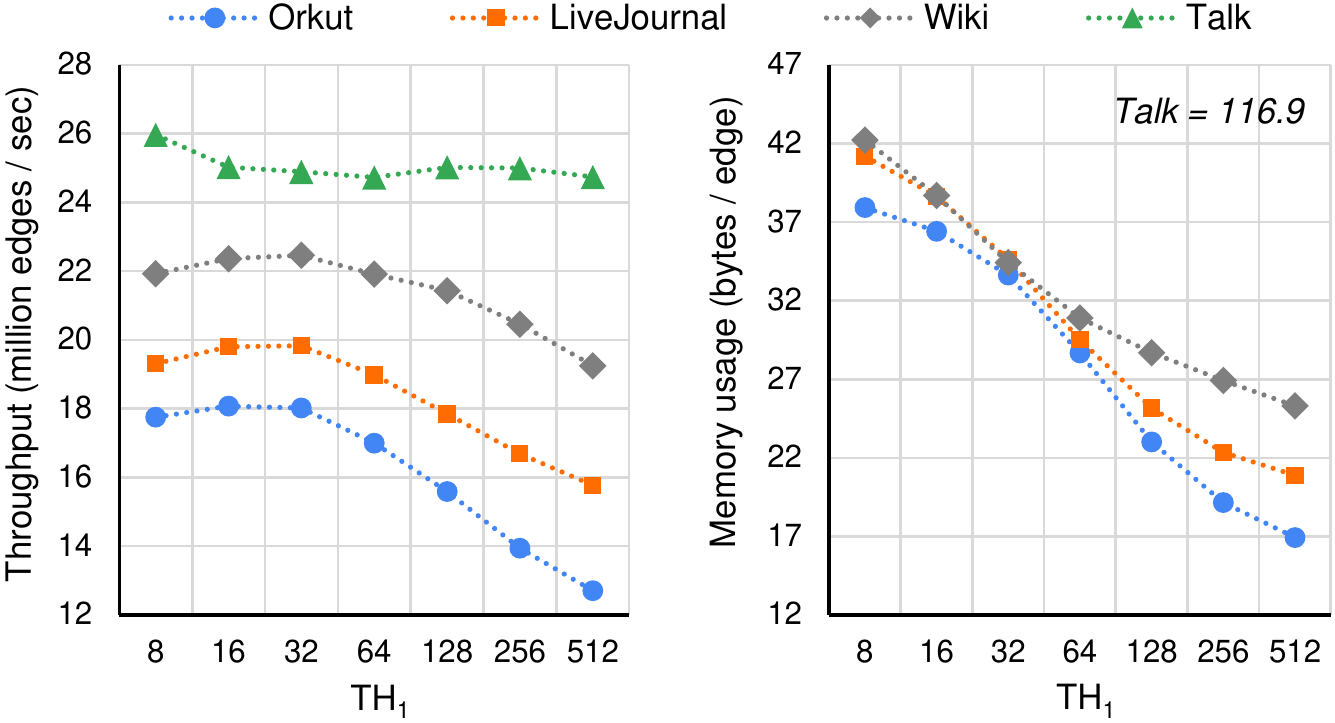}
	\caption{Impact of $TH_1$ threshold on update throughput and memory usage.}
	\label{fig:th1_sweep}
\end{figure}

\subsection{Impact of $TH_1$ Threshold}
\label{ssec:th1}

Figure \ref{fig:th1_sweep} shows the impact of the $TH_1$ threshold on update throughput and memory usage. Analytics throughput is not shown because the choice of $TH_1$ does not impact the analytics performance. $TH_1$ of 16 and 32 provides the best throughput on three out of the four datasets. We used $TH_1$ of 32 in all other experiments, as it has lower memory usage.

The $TH_1$ threshold controls the ratio between Type2 and Type3 vertices. Increasing $TH_1$ maps more higher-degree vertices to Type2 instead of Type3. As Type2 vertices requires linear search during updates, it eventually becomes a performance bottleneck for $TH_1 > 32$. On the other hand, Type2 vertices do not require maintaining a hash table and thus require less memory than Type3. As a result, increasing $TH_1$ reduces the memory usage. On average, increasing $TH_1$ from 8 to 512 reduces the memory usage by 1.9x.

The Talk dataset is an outlier showing negligible variation with $TH_1$. This is because, for the Talk dataset, 98.3\% of the vertices are mapped to Type1 (refer to Table \ref{tab:datasets}), leaving only 1.7\% of the vertices that can be affected by changing $TH_1$.

\subsection{Impact of Optimizations}
\label{ssec:opt}

\begin{table}[t]
    \centering
    \caption{Impact of Optimizations on the Update Throughput (Baseline is the proposed hybrid format without any optimizations applied)}
    \label{tab:opt}
    \footnotesize
    \renewcommand{\arraystretch}{1.3}
    \setlength\tabcolsep{4.7pt}
    \begin{threeparttable}
        \begin{tabular}{l|c|c|c|c|c}
            \toprule
            \multirow{2}{*}{Configuration} &\multirow{2}{*}{Format} & Allocation & Hashing &\multicolumn{2}{c}{Speedup} \\
            & & Scheme & Scheme & STail & HTail \\
            \midrule
            baseline & Hybrid &malloc\tnote{2} &std\_map\tnote{3} &1.00 &1.00 \\
            next best\tnote{1} &- &- &- &0.76 &0.29 \\
            opt pool & Hybrid &proposed &std\_map\tnote{3} &1.12 &1.14 \\
            opt hash & Hybrid &malloc\tnote{2} &proposed &1.71 &1.70 \\
            \hline
            GT\_Tessil & Hybrid &proposed & Tessil\tnote{4} &1.40 &1.60 \\
            GT\_RHH & Hybrid &proposed & RHH\tnote{5} &1.35 &1.45 \\
            GT\_Abseil & Hybrid &proposed &Abseil\tnote{6} &1.34 &1.43 \\
            \textbf{GraphTango} &\textbf{\textit{Hybrid}} & \textbf{\textit{proposed}} &\textbf{\textit{proposed}} &\textbf{1.89} &\textbf{1.79} \\
            \hline
            DegAwareRHH &DegAware & malloc\tnote{2} & RHH &0.29 &0.29 \\
            DegAwareCFH &DegAware & malloc\tnote{2} &proposed &0.40 &0.38 \\
            \bottomrule
        \end{tabular}
        \begin{tablenotes}
            \scriptsize
            \item[1] AdListShared for STail and DegAwareRHH for HTail.
            \item[2] glibc version 2.31.
            \item[3] std::unordered\_map with libstdc++ version 6.0.28.
            \item[4] tsl::robin\_map from \cite{tessil}, version 1.0.1.
            \item[5] Robin Hood hashing implementation from \cite{rhhgit}, version 3.11.5.
            \item[6] Google's Abseil flat\_hash\_map version LTS 20211102
            \cite{abseilgit}.
            
        \end{tablenotes}
\end{threeparttable}
\vspace{0.2cm}
\end{table}

\begin{figure*}[t]
	\centering
	\includegraphics[clip, trim={0.1cm 4.6cm 1.2cm 0.0cm}, scale=0.56]{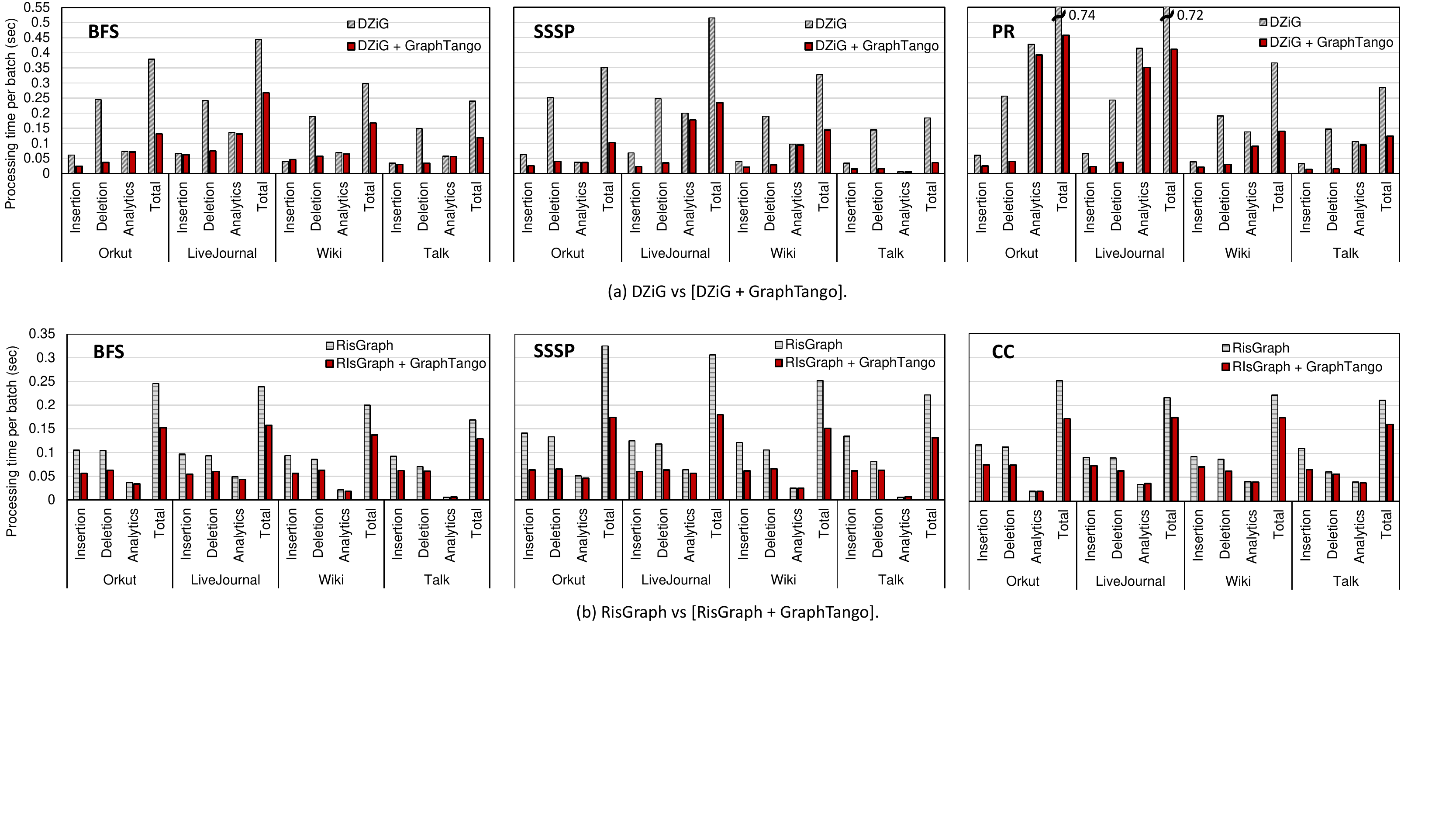}
	\caption{Batch processing time breakdown of DZiG and RisGraph integration. \textit{Lower is better}.}
	\label{fig:dzig_ris}
	\vspace{0.2cm}
\end{figure*}

The purpose of this section is to isolate the contribution of the proposed hybrid format as well as the memory allocation and hashing scheme optimizations. Table \ref{tab:opt} shows our findings. In this table, the STail and HTail columns show the normalized update throughput over the \textit{baseline} configuration for short-tailed and heavy-tailed graphs, respectively. We show only the update throughput because the memory pool and hashing optimizations have a negligible impact on the analytics throughput. 

The \textit{baseline} configuration implements our proposed hybrid format (i.e., Type1, Type2, and Type3 mapping of vertices) but uses sub-optimal malloc for memory allocation and std::unordered\_map for hashing. We can observe that using only the hybrid format is sufficient to provide a better performance than the \textit{next best} approach (AdListShared for STail and DegAwareRHH for HTail). Compared to the next best approach, the \textit{baseline} offers 1.3x (3.4x) higher speedup for STail (HTail) graphs. The \textit{opt pool} configuration shows the benefit of the proposed memory allocation scheme. On average, the proposed pool provides 1.13x better performance over the \textit{baseline}. On the other hand, the \textit{opt hash} configuration shows the benefit of the proposed cache-friendly hashing scheme, offering 1.71x speedup over std::unordered\_map. With both optimizations enabled, \gname{} provides an average speedup of 1.84x over the \textit{baseline}.

A valid concern at this point is whether we can combine other hashing schemes with our hybrid format to get even better performance. To answer this question, we tried three other open-addressing-based hash-table implementations: i) \textit{GT\_Tessil}: The Robin Hood hashing variation of Tessil (tsl::robin\_map) \cite{tessil}. This is the fastest hash table implementation according to the benchmark results published in \cite{rhhdoc}. ii) \textit{GT\_RHH}: Another fast implementation of Robin Hood hashing from \cite{rhhgit, rhhdoc}. Although it is slightly slower than Tessil, it consumes significantly less memory. iii) \textit{GT\_Abseil}: Google's Abseil flat\_hash\_map \cite{abseilgit, kulukundis2017designing}. The max load factors of these approaches are set equal to ours $(=0.5)$ for a fair comparison. On STail (HTail) graphs, the proposed hashing scheme provides 1.35x (1.12x) speedup over \textit{GT\_Tessil}, 1.4x (1.23x) speedup over \textit{GT\_RHH} and 1.41x (1.25x) speedup over \textit{GT\_Abseil}. Because our hashing scheme tries to minimize cache line access, it is especially suitable for graph workloads where the hash table is unlikely to reside in the cache.

Finally, we evaluate whether DegAwareRHH can leverage our hashing scheme to outperform \gname{}. \textit{DegAwareCFH} denotes this configuration. \textit{DegAwareCFH} provides 1.37x (1.31x) better throughput over the vanilla DegAwareRHH. However, \gname{} still outperforms it by 4.7x for both STail and HTail graphs.

\subsection{Integration with DZiG and RisGraph}
\label{ssec:dzig}

This section demonstrates that full-fledged graph processing frameworks can leverage the \gname{} format to improve their performance further. We selected two state-of-the-art graph processing frameworks DZiG \cite{mariappan2021dzig} and RisGraph \cite{feng2021risgraph} for this purpose. We modified their publicly available source code \cite{dzig_github,risgraph_code} and replaced their storage format with \gname{}. We run the datasets on BFS, PR, and SSSP for DZiG. CC is omitted because its implementation is unavailable in the framework's repository. For the same reason, PR is omitted in case of RisGraph.

Figure \ref{fig:dzig_ris}(a) shows the comparison results between DZiG and DZiG+\gname{}. DZiG internally uses adjacency list as graph storage. For this reason, analytics time for DZiG and DZiG+\gname{} is similar in most cases. Interestingly, the insertion time is also comparable in some cases. For example, LiveJournal and Wiki datasets for BFS. This is because the original DZiG's edge insertion does not check for duplicate edges. Therefore, the edge insertion becomes as simple as adding an element to the end position of an array \footnote{There is a flag to enable duplicate edge insertion checking. But that checking is done by sorting the batch as a pre-processing step, thereby incurring heavy overhead.}. On average, \gname{} provides a 1.9x reduction in insertion time {\em even though it also checks for duplicate edges}. For deletion, unmodified DZiG performs 6x worse on average. We identified two reasons: i) Unlike insertions in DZiG that do not search for duplicates, delete operations require a linear search through the neighbor list, incurring higher runtime cost, and ii) DZiG performs a quicksort on the batch based on the source and destination vertex ids to distribute them among the threads. As we use fixed mapping of vertices in \gname{}, sorting costs are avoided. Overall, DZiG+\gname{} provides an average of 2.3x reduction in total batch processing time compared to the original DZiG.

Figure \ref{fig:dzig_ris}(b) shows the comparison between RisGraph and RisGraph+\gname{}. RisGraph uses a hybrid graph storage format that uses adjacency list for low/medium degree vertices and adjacency list along with hash table for high degree vertices. Unlike GraphTango, RisGraph does not differentiate between low and medium degree vertices and uses the same data structure for both. Furthermore, RisGraph uses Google's dense hash map and does not attempt to minimize the number of cache accesses as GraphTango does with its proposed cache-friendly hashing scheme. Due to these differences, RisGraph+\gname{} provides on average 1.5x reduction in total batch processing time compared to the vanilla RisGraph.

\section{Conclusions}

Existing streaming graph representation formats can only support either short-tailed or heavy-tailed workloads efficiently. This paper proposes \gname{}, which aims to solve this issue by adaptively switching formats based on the current degree of a vertex. We also propose a cache-efficient hashing scheme and a fast memory pool. These optimizations work in synergy with \gname{} to provide excellent update and analytics throughput regardless of the graph's degree distribution. Our evaluation on the SAGA-Bench showed that on average (maximum), \gname{} provides 4.5x (6.6x) higher insertion throughput, 3.2x (5.0x) higher deletion throughput, and 1.1x (1.6x) higher analytics throughput over the \textit{next best} approach.

\bibliographystyle{unsrt}
\bibliography{reference}

\end{document}